\documentclass{ecai}

\usepackage{times}
\usepackage{latexsym}
\usepackage{graphicx}  %Required
\usepackage{color}
\usepackage{multirow}
\usepackage{xspace}
\usepackage{subfig}
\usepackage{amssymb}
\usepackage{amsmath}
\usepackage[normalem]{ulem}
\usepackage[table]{xcolor}
\usepackage{caption}

\ecaisubmission

\definecolor{lightyellow}{rgb}{1.0, 1.0, 0.88}
\definecolor{lightcyan}{rgb}{0.88, 1.0, 1.0}

\newcommand{\ps}{\textsf{PS}}
\newcommand{\mix}{\textsf{MIX}}
\newcommand{\swap}{\textsf{SWAP}}
\newcommand{\move}{\textsf{MOVE}}
\newcommand{\swapps}{\textsf{SWAP-PS}}
\newcommand{\swapmix}{\textsf{SWAP-MIX}}
\newcommand{\swapmixaswap}{\textsf{SWAP-MIX-AS-SWAP}}
\newcommand{\init}{\textsf{INIT}}
\newcommand{\doneps}{\textsf{DonePS}}

\begin{document}

\title{Planning for Compilation of a Quantum Algorithm \\ for Graph Coloring}

\author{Minh Do\institute{KBR \& NASA ARC, USA, minh.do@nasa.gov} 
        \and Zhihui Wang\institute{USRA \& NASA ARC, USA, zhihui.wang@nasa.gov}
        \and Bryan O'Gorman\institute{UC Berkeley \& NASA ARC, USA, bogorman@berkeley.edu} \\
        \and Davide Venturelli\institute{USRA \& NASA ARC, USA, davide.venturelli@nasa.gov}
        \and Eleanor Rieffel\institute{NASA ARC, USA, eleanor.rieffel@nasa.gov} 
        \and Jeremy Frank\institute{NASA ARC, USA, Jeremy.D.Frank@nasa.gov}}

\maketitle
\bibliographystyle{ecai}

\begin{abstract}
The problem of compiling general quantum algorithms for implementation on near-term quantum processors has been introduced to the AI community. Previous work demonstrated that temporal planning is an attractive approach for part of this compilation task, specifically, the routing of circuits that implement the Quantum Alternating Operator Ansatz (QAOA) applied to the MaxCut problem on a  quantum processor architecture.
 In this paper, we extend the earlier work to route circuits that implement QAOA for Graph Coloring problems.
QAOA for coloring requires execution of more, and more complex, operations on the chip, which makes routing a more challenging problem. We evaluate the approach on state-of-the-art hardware architectures from leading quantum computing companies. Additionally, we apply a planning approach to qubit initialization. Our empirical evaluation shows that temporal planning compares well to reasonable analytic upper bounds, and that solving qubit initialization with a classical planner generally helps temporal planners in finding shorter-makespan compilations for QAOA for Graph Coloring. These advances suggest that temporal planning can be an effective approach for more complex quantum computing algorithms and architectures.
\end{abstract}

%----------
\section{Introduction}
\label{sec:intro}

Quantum computers apply quantum operations, called quantum gates, to qubits, the basic memory unit of quantum processors. Quantum algorithms are often specified as quantum circuits on idealized hardware, in which perfect gates can be applied to any set of qubits, whereas physical hardware has various constraints and imperfections. In practice, these idealized quantum circuits must be compiled to specific hardware. One common way of overcoming the restricted connectivity of such hardware is by adding additional gates that route qubit states to locations where the desired gate can act on them. Compilations that minimize the overall execution duration return results more quickly. More importantly, decoherence effects can destroy the computation in a short time and thus minimizing computation time is therefore vital to obtain results on near-term quantum hardware that does not support significant quantum error correction. 

Recently, the use of temporal planners to compile quantum circuits was explored for QAOA applied to the MaxCut problem~\cite{vent:ijcai17}; machine operations were modeled as PDDL2.1 durative actions, enabling domain-independent temporal planners to find a parallel sequence of conflict-free operations to implement the high-level quantum algorithm.
Several state-of-the-art temporal planners were used to show empirically that temporal planning is a promising approach to compile circuits of various sizes to a model hardware chip featuring the essential characteristics of newly emerging quantum hardware.
Building upon this work, several subsequent works have utilized different techniques to more effectively solve the same routing instance set.
In~\cite{booth:icaps18}, the authors extended the planning-based approach by integrating it with a constraint-programming solver to further improve the plan quality. Greedy randomized search and genetic algorithms
are explored in ~\cite{oddi:cpaior18,rasconi:aaai19};
while those approaches provide improved results, they require building domain-dependent heuristics or encodings that may lack the flexibility of the model-based domain-independent temporal-planning approach introduced in~\cite{vent:ijcai17}, which can work on different classes of routing instances with a variety of hardware constraints and configurations.

In this paper, we expand the scope of the routing problem with a new target domain: QAOA for Graph Coloring~\cite{hadfield2019quantum,wang2019xy}; specifically, the optimization variant in which the number of properly colored edges is maximized. Compiling QAOA for Graph Coloring is different, harder, and more general than compilation of QAOA for MaxCut because of: (1) the existence of mix operations on two logical qubits (qstates); this leads to much more contention for resources on the gate-model hardware; and (2) the compilation task itself is more complex than it is for MaxCut. Thus, solving this problem class is key to future efforts to effectively utilize real-world gate-model quantum computers.
Our main contributions over previous work are:

\vspace{-1mm}

\begin{itemize}
    \item \textit{New instance class}: while our previous work concentrated on routing of QAOA for MaxCut, here we investigate routing of QAOA for Graph Coloring. Compiling Graph Coloring into a physical circuit requires a different set of gates, which include `hybrid' gates, and more complex ordering between them, making the compilation task much more complicated. To our knowledge, we present a compilation study on the most complex Noisy-Intermediate Scale Quantum (NISQ) optimization algorithm application to date. 
    \item \textit{More diverse physical hardware architectures}: while previous work used an earlier hypothetical model from Rigetti, the computational results for this paper were conducted using hardware graphs and gate durations that are closer to the real hardware that Google, IBM, and Rigetti are building.
    \item \textit{Qubit initialization}: we present initial research results singling out the problem of qubit initialization (QI), which is a sub-problem related to routing. Our approach of using classical planning to solve QI improves the performance of all tested temporal planners across a variety of problem setups.
\end{itemize}

The paper is structured as follows. The next section provides background on quantum circuit routing. Then, in Section~\ref{sec:qcc_graphcoloring}, we outline the problem of circuit routing for QAOA on Graph Coloring. Section~\ref{sec:qcc_temporalplanning} describes how it can be modeled as a temporal planning problem using the PDDL2.1 standard modeling language. Section~\ref{sec:initialization} describes our approach to using planning to initialize the qstates in order to minimize makespan. In Section~\ref{sec:evaluation}, we describe different experiments and results showing the viability of our approach. Section~\ref{sec:conclusion} concludes the paper and outlines some of our future work directions.

%----------
\section{Quantum Circuit Routing}
\label{sec:qcc}

General quantum algorithms historically have been described in an idealized architecture in which a gate can act on any subset of qubits. However, in an actual superconducting qubit device, such as the latest chips manufactured by IBM, Rigetti Computing, Google and  Intel~\cite{corcoles2019challenges, murali2019full, arute2019quantum}, physical constraints impose restrictions on the sets of qubits on which gates can be performed. Recently, a significant number of approaches have been explored for compiling idealized quantum circuits to realistic quantum hardware with a specific focus on ``circuit routing''\footnote{Previous work referred to this as ``quantum circuit compilation'' (QCC).} (i.e., swap gate insertion strategies) in NISQ devices, targeting algorithms that could be run in the near-term~\cite{cowtan2019qubit, zulehner2018efficient, li2019tackling, rasconi:aaai19, oddi:cpaior18, gokhale2019partial, itoko2019optimization}.

For superconducting qubit architectures, qubits in these quantum processors can be thought of as nodes in a planar graph, with 2-qubit quantum gates associated with edges and 1-qubit quantum gates associated with nodes.  Gates that operate on distinct sets of qubits may be able to operate concurrently, subject to additional restrictions, such as requiring the sets involved with concurrent gates to be non-adjacent. Furthermore, there are different types of quantum gate, each taking different duration that is dependent on the specific physical implementation. In order for the computation specified by the idealized circuit to be carried out, a particular type of 2-qubit gate, the {\textit{swap}} gate, is often applied to exchange the state of two qubits. A sequence of swap gates moves the logical states of two distant qubits to a location where a desired gate can be applied. Swap gates may be available only on a subset of edges in the hardware graph, and swap duration may depend on where they are located. In this paper, we will consider the case in which swap gates are available between any two adjacent qubits on the chip, and all swap gates have the same duration; the more general cases are a straightforward generalization.

%------
\begin{figure}[tb]
  \centering
  \includegraphics[width=0.9\columnwidth]{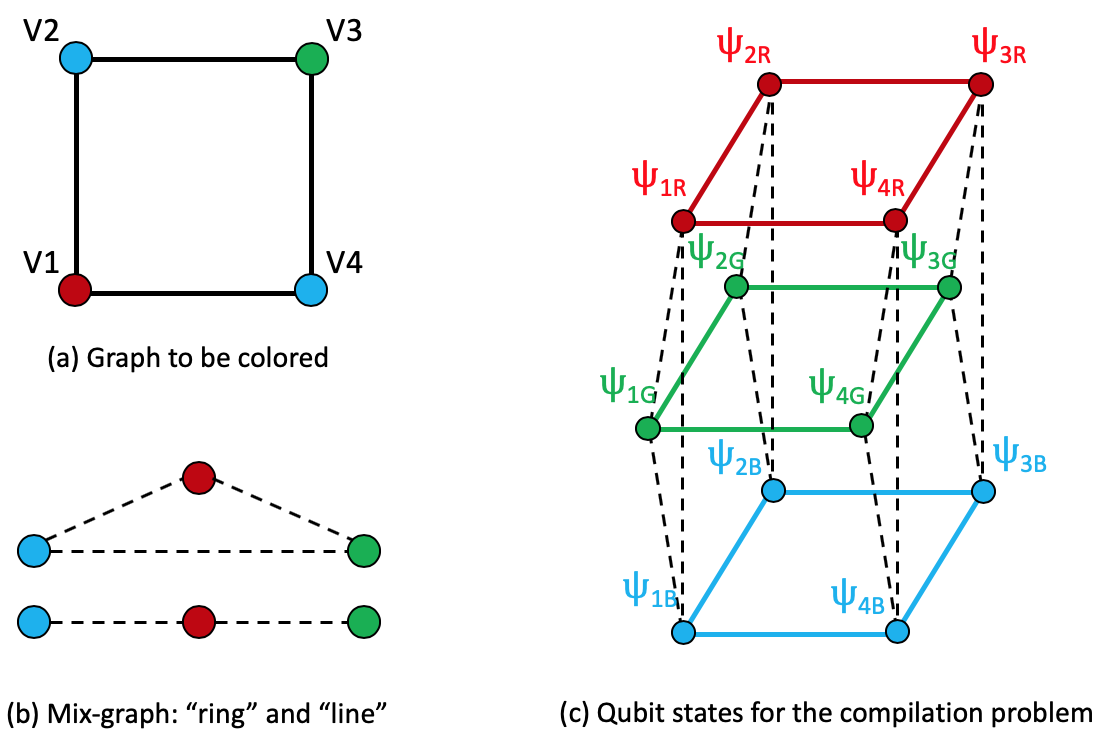}
  \caption{Routing for Graph Coloring: coloring a square with 3 colors.}
    \label{fig:example}
\end{figure}
%-------

%----------
\section{Circuit Routing for Graph Coloring}
\label{sec:qcc_graphcoloring}

In this paper, the Graph Coloring problem we will investigate is the \textit{vertex coloring} problem in which all $n$ vertices $v \in V$ of a given graph $\mathcal{G} = \{V,E\}$  are colored with $k$ colors. The objective function is to maximize the number of edges $e = \langle v_1, v_2 \rangle \in E$ where vertices $v_1$ and $v_2$ are colored differently. This problem exemplifies the combinatorial structure of many scheduling and asset-allocation problems in industry and computer science research. Figure~\ref{fig:example}(a) shows a concrete example in which a square is properly colored with 3 colors. 

The quantum algorithm that we follow is a variant of the ``Quantum Alternating Operator Ansatz''~\cite{hadfield2019quantum}, a generalization of the ``Quantum Approximate Optimization Algorithm'' (QAOA)~\cite{Farhi14}, applied to the Graph Coloring problem.

\vspace{1mm}

\noindent {\bf The Ideal Circuit:} the idealized QAOA circuit for Graph Coloring has been studied in~\cite{hadfield2019quantum} and~\cite{wang2019xy}. We refer the quantum-computing skilled reader to those references for understanding the algorithm, and focus here only on its compiler-level implementation. It is specified by two types of 2-qubit gate, the \textit{phase separation} (\ps) gate and the \mix\ gate, which need to be applied to a  set of instance-specific problem \textit{goals}\footnote{Additional single qubit gates are also required in graph coloring but they are not relevant for routing since they can be executed at durations negligible compared to two-qubit gates, so we will disregard them in this paper.}. Each goal specifies a pair of \textit{qubit states} (qstate), the information content of a qubit, that must have a \ps\ or \mix\ ``goal'' gate applied to them. For the Graph Coloring problem, each qstate represents a \textit{vertex-color} combination. Thus, for our leading example problem of coloring a graph of 4 vertices ($V = \{V_1,V_2,V_3,V_4\}$) with 3 colors (red [R], green [G], and blue [B]), there are a total of $4 \times 3 = 12$ qstates: $\Psi = \{\psi_{1R},\psi_{2R},\ldots,\psi_{4B}\}$, illustrated in Figure~\ref{fig:example}(c). The idealized Graph Coloring circuit is specified as follows:

%------
\begin{figure}[tb]
\centering
\includegraphics[width=\columnwidth]{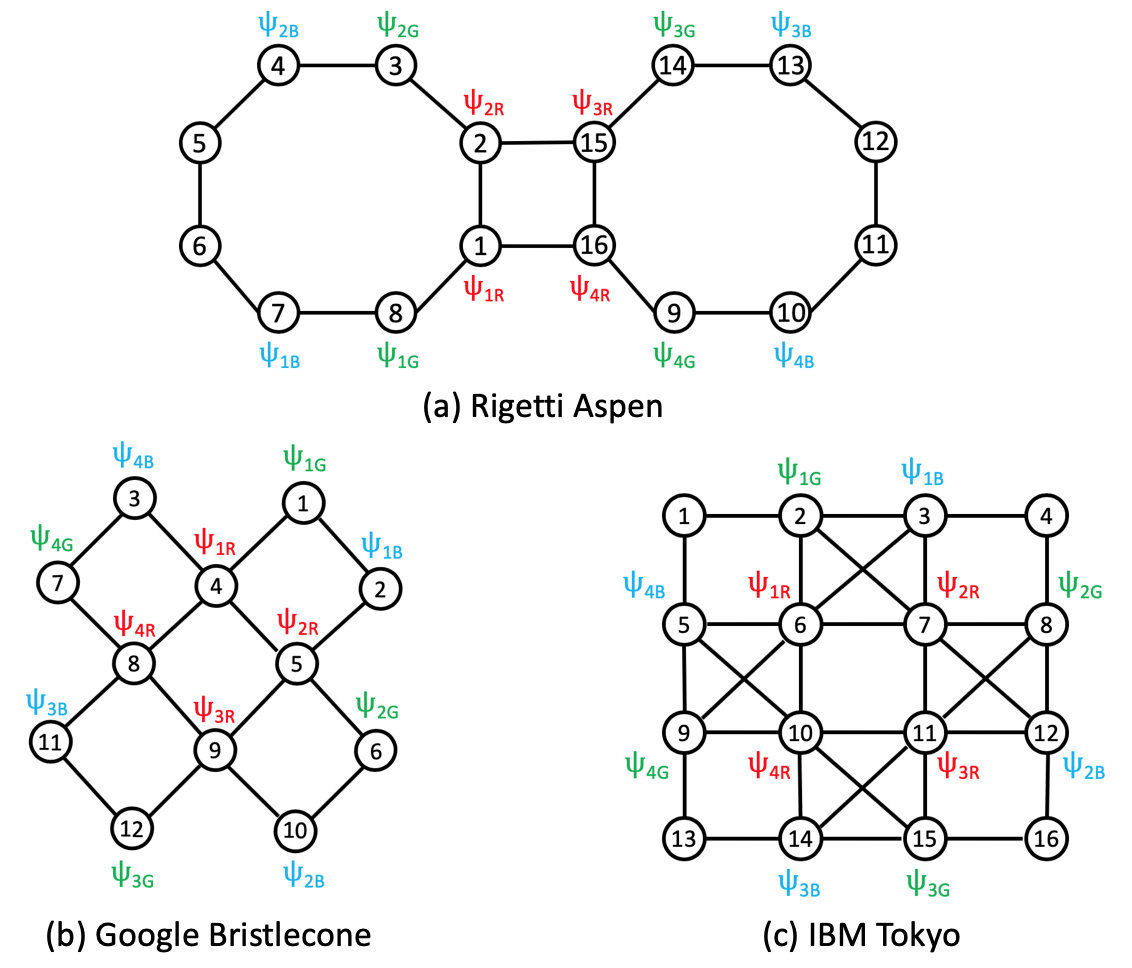}
\caption{Example hardware chip layout from different companies.}
\label{fig:machine}
\end{figure}
%-------

\begin{itemize}
    \item \textit{\ps\ gate requirements}: for QAOA for Graph Coloring, there should be one \ps\ gate between any pair of qstates $\psi_{iC}$ and $\psi_{jC}$ that: (1) represent the same color $C$; (2) participate in an edge $e = \langle V_i, V_j \rangle \in E$.
    We denote the application of a \ps\ gate as, e.g., \ps($\psi_{1R},\psi_{2R}$), \ps($\psi_{1R},\psi_{4R}$), \ps($\psi_{3B},\psi_{4B}$). 
    
    \item \textit{\mix\ gate requirements}: one parameter of the QAOA for Graph Coloring is the ``mix-graph'' $\mathcal{G}_{mix}$ which consists of: (1) nodes representing different colors; and (2) edges representing which pairs of colors require \mix\ gates. In essence, mixing operators generate transitions between states representing different colors of the same vertex. Efficient mixing plays an important role in the quantum computation by achieving  constructive interference that leads to a good solution. A complete mixing-graph would allow all colors to transit to each other faster than a minimally-connected mixing graph (a chain) where a color only transits to its neighboring colors in one step~\cite{wang2019xy}. On the other hand, a denser mixing-graph will require more \swap\ gates, leading to a longer circuit execution time. Figure~\ref{fig:example}(b) shows two examples of a mix-graph for the three colors used in our example: (1) ``ring'': each color is connected to the other two. If we remove one edge (e.g., $Green-Blue$), then we have a  (2) ``line'' mix-graph. Given a mix-graph $\mathcal{G}_{mix}$, the requirement is to achieve one \mix\ gate for each pair of qstates $\psi_{iC_j}$ and $\psi_{iC_k}$ that: (1) represent the same vertex $V_i$; (2) are associated with two different colors $C_j$ and $C_k$ such that $\langle C_j, C_k \rangle$ is an edge in $\mathcal{G}_{mix}$. We denote the application of a \mix\ gate as, e.g, \mix($\psi_{1R},\psi_{1G}$), \mix($\psi_{1R},\psi_{1B}$), \ldots, \mix($\psi_{4G},\psi_{4B}$).

    \item \textit{Goal orderings}: constraints on the circuit structure of QAOA enforce orderings between the \ps\ and \mix\ gates:  all \ps\ gates involving a certain qstate $\psi_{iC_j}$ should be completed before any \mix\ gate involving $\psi_{iC_j}$ can be executed. Let's take the qstate $\psi_{1R}$ as an example: the two \ps\ gates \ps($\psi_{1R},\psi_{2R})$ and \ps($\psi_{1R},\psi_{4R})$ need to be completed before either of the two \mix\ gates:
    \mix($\psi_{1R},\psi_{1G})$ or \mix($\psi_{1R},\psi_{1B})$ can start.
    Note that goal ordering constraints do not enforce that {\em all} \ps\ gates need to be completed before the \mix\ gates can start.
\end{itemize}

\noindent {\bf Architecture-specific operations:} as described in Section~\ref{sec:qcc}, while the idealized quantum circuit for Graph Coloring contains the set of goal \ps\ and \mix\ gates along with their orderings and the assumption that any goal gate can be applied anytime, they need to be ``compiled'' into architecture-specific operations that can actually be executed on a particular quantum chip that has numerous hardware constraints. Figure~\ref{fig:machine} shows several examples: (a) a 16-qubit Aspen chip layout by Rigetti Computing; (b) a 12-qubit section from a 72-qubit Bristlecone chip by Google; and (c) a 16-qubit section of IBM's 20-qubit Tokyo architecture. 

The set of operations for Graph Coloring is significantly more complicated compared to the previous work on planning for routing of MaxCut~\cite{vent:ijcai17,booth:icaps18} where there are only three operations: 2-qubit \ps\ operation, 1-qubit \mix\ operation, and 2-qubit \swap\ operation. The increase in complexity arises also due to the availability of `multi-purpose' or `hybrid' operations that accomplish multiple objectives (e.g., \swap\ + \mix). More specifically, the following types of operation need to be considered by the compiler to tackle QAOA for Graph Coloring on those architectures:

\vspace{-0.5mm}

\begin{itemize}
    \item \textit{\ps\ and \mix\ operations}: direct implementations of the ideal circuit \ps\ and \mix\ gates that can be in principle applied to any pair of qstates residing on two qubits that are adjacent/interconnected on the hardware chip.
    
    \item \textit{\swap\ operation:} swaps the locations of two qstates residing on two interconnected qubits. For example, taking the layout of qstates on the Rigetti chip (Figure~\ref{fig:machine}(a)): \swap($\langle \psi_{2B},q_4 \rangle, \langle \psi_{2G},q_3 \rangle)$ leads to: $\langle \psi_{2B}, q_3 \rangle$ and $\langle \psi_{2G}, q_4 \rangle$.
    
    \item \textit{\move\ operation}: a variant of the \swap\ operation that instead of swapping the locations of the two adjacent qstates, it moves a qstate to an adjacent empty qubit. For example (Figure~\ref{fig:machine}(a)): \move($\langle \psi_{2B},q_4 \rangle, q_5$) leads to $\langle \psi_{2B}, q_5 \rangle$, leaving $q_4$ empty\footnote{This operation does correspond to a \swap\ operation in the real chip, where empty qubits don't exist. It is defined only for modeling convenience.}.
    
    \item \textit{\swapps\ operation}: this operation combines the effects of the \swap\ and \ps\ operations. Thus, \swapps($\langle \psi_{1R},q_1 \rangle, \langle \psi_{2R},q_2 \rangle)$ in Figure~\ref{fig:machine}(a) will switch the locations of $\psi_{1R}$ and $\psi_{2R}$ like the \swap\ operation but also accomplish \ps($\psi_{1R}, \psi_{2R}$). For a pair of qstates that do not have a \ps\ goal gate requirement,  the \swapps\ gate can be applied, but its effect is identical to using a \swap\ gate. Thus:  \swapps($\langle \psi_{2B},q_4 \rangle, \langle \psi_{2G},q_3 \rangle$) has the same effect as \swap($\langle \psi_{2B},q_4 \rangle, \langle \psi_{2G},q_3 \rangle)$ since there is no goal gate requirement \ps($\psi_{2B}, \psi_{2G}$). Since the \swap\ and \swapps\ operations may have different durations depending on the particular physical connection, the planner needs to decide whether to use one over the other at a particular location on the chip. 
    
    \item \textit{\swapmix\ operation}: similar to the \swapps\ operation, this one combines the effects of the \swap\ and the \mix\ operations. However, unlike \swapps\ that can be applied to any pair of qstates on adjacent qubits, the \swapmix\ operation can only be applied between qstates representing nodes in the same mix-graph (Figure~\ref{fig:example}(b)). Let's assume the ``line" mix-graph in Figure~\ref{fig:example}(b), which has two connections $B \leftrightarrow R$ and $R \leftrightarrow G$, then: (1) \swapmix($\langle \psi_{1R},q_1 \rangle, \langle \psi_{1G},q_8 \rangle$) has the combined effects of \swap\ and \mix\ gates; (2) \swapmix($\langle \psi_{1B},q_7 \rangle, \langle \psi_{1G},q_8 \rangle$) has the same effect as \swap\ gate (but can have a shorter makespan to make it a better option than \swap) since $\psi_{1B}$ and $\psi_{1G}$ are not connected on, but belong to the same, the mixgraph; (3) \swapmix($\langle \psi_{1R},q_1 \rangle, \langle \psi_{2R},q_R \rangle$) are not allowed since they are not connected on a mix-graph.
\end{itemize}

\vspace{-1mm}

\noindent {\bf Problem definition:} Given an idealized circuit, comprised of \ps\ and \mix\ gates, used  to define a QAOA quantum algorithm for Graph Coloring, the circuit routing problem  is  to  find  a  new  architecture-specific  circuit  that implements  the  idealized  quantum  circuit, utilizing the architecture-specific \ps, \mix, \swap, \move, \swapps, and \swapmix\ operations as required. The objective is to minimize the overall duration to execute all operations in the new circuit.

%----------
\section{Model Circuit Routing for Graph Coloring as a Temporal Planning Problem}
\label{sec:qcc_temporalplanning}

Planning is the problem of finding a conflict-free set of actions and their respective execution times that connects the \emph{initial-state} $I$ and the desired \emph{goal state} $G$. We now introduce some key background concepts for the routing-as-temporal planning problem.

\vspace{1mm}

\noindent \emph{Planner:} a planner takes as input a specification of domain and problem instance, and returns a valid plan, if one exists. At the abstract level, the planner needs to solve the QAOA compilation problem described in the previous section: taking as input the required \ps\ and \mix\ gates and utilizing the architecture and problem-specific operations to build a plan achieving all those gates.

\vspace{1mm}

\noindent \emph{Planning Domain Description Language (PDDL):} The de-facto standard modeling languages used by many domain-independent planners. We use PDDL 2.1~\cite{pddl21}, which allows the modeling of temporal planning formulations in which every action $a$ has duration $d_a$, starting time $s_a$, and end time $e_a = s_a + d_a$. Action conditions are required to be satisfied either (1) instantaneously at $s_a$ or $e_a$ or (2) to be true starting at $s_a$ and remain true until $e_a$. Action effects may instantaneously occur at either $s_a$ or $e_a$. Actions can execute when their temporally-constrained conditions are satisfied; and when executed will cause state-change effects. The most common objective function in temporal planning is to minimize the plan \emph{makespan}, i.e., the shortest total plan execution time. This objective matches well with the objective of our targeted routing problem: minimizing the total circuit execution time (i.e., circuit depth).\\

%------
\begin{figure}[t]
  \centering
  \begin{minipage}{0.5\textwidth} 
  \footnotesize{
  (:durative-action swap\_mix\_at\_q1\_q2 \\
   \hspace*{0.5cm}:parameters (?s1 - qstate ?s2 - qstate) \\
   \hspace*{0.5cm}:duration (= ?duration 1.0) \\
   \hspace*{0.5cm}:condition (and (at start (located\_at\_q1 ?s1)) \\
   \hspace*{2.2cm}     	 (at start (located\_at\_q2 ?s2)) \\
   \hspace*{2.2cm}     (at start (ps\_completed ?s1)) \\
   \hspace*{2.2cm}     (at start (ps\_completed ?s2)) \\
   \hspace*{2.2cm}    (at start (mixgraph\_edge ?s1 ?s2)) \\
   \hspace*{2.2cm}     (at start (not (mixed ?s1 ?s2)))) \\
   \hspace*{0.5cm}:effect (and (at start (not (located\_at\_q1 ?s1))) \\
   \hspace*{1.8cm}         (at start (not (located\_at\_q2 ?s2))) \\
   \hspace*{1.8cm}         (at end (located\_at\_q1 ?s2)) \\
   \hspace*{1.8cm}         (at end (located\_at\_q2 ?s1)) \\
   \hspace*{1.8cm}         (at end (mixed ?s1 ?s2)) \\
   \hspace*{1.8cm}        (at end (mixed ?s2 ?s1))))
   }  
  \end{minipage}  
  \caption{ PDDL model of the \swapmix\ operation.}
%  \vspace{-0.1in}
  \label{fig:PDDL}
\end{figure}
%-------

\noindent{\bf Modeling Routing for Graph Coloring in PDDL 2.1:} Following the software structure of the end-to-end compiler tool-chain presented in~\cite{venturelli2019quantum}, at the highest level, we need to take as input:

\vspace{-1.0mm}
\begin{itemize}
 \item Qstates and their relations: this in turn encapsulates the graph to be colored $\mathcal{G}$ (Figure~\ref{fig:example}(a)) and the ``mix-graph'' $\mathcal{G}_{mix}$ (Figure~\ref{fig:example}(b)).
 \item The $\mathcal{G}_{machine}$ graph representing the hardware chip layout (Figure~\ref{fig:machine}).
\end{itemize}
\vspace{-1.0mm}

\noindent and turn them into \emph{objects}, \emph{predicates}, \emph{actions}, \emph{initial state}, and \emph{goal state} of a temporal planning problem such that a valid plan represents a parallel sequence of architecture-specific operations (see Section~\ref{sec:qcc_graphcoloring}) enabling all required \ps\ and \mix\ gates.

\vspace{1mm}

\noindent \emph{Objects:} as in~\cite{vent:ijcai17}, we model qstates as PDDL objects. Physical qubits and the connections in $\mathcal{G}_{machine}$, $\mathcal{G}_{mix}$, and $\mathcal{G}$ graphs are modeled implicitly with the list of predicates and action descriptions.

\vspace{1mm}

\noindent \emph{Predicates:} the following facts are represented by PDDL predicates:

\vspace{-2mm}
\begin{itemize}
    \item $\textup{located\_at\_}q(s)$: if a qstate $s$ is located at a given qubit $q$.
    \item $\textup{empty\_}q$: if a given qubit $q$ is empty. This predicate enables the \move\ action (see Section~\ref{sec:qcc_graphcoloring}).
    \item $\textup{psed}(s_1,s_2)$ and $\textup{mixed}(s_1,s_2)$: if a \ps\ or \mix\ gate has been accomplished between a pair of qstates $s_1$ and $s_2$.
    \item $\textup{edge}(s_1,s_2)$: if two qstates $s_1$ and $s_2$ are connected in the graph $\mathcal{G}$ to be colored. This predicate serves as a precondition of the \ps\ action.
    \item $\textup{mixgraph\_edge}(s_1,s_2)$: if qstates $s_1$ and $s_2$ are connected in the mix-graph $\mathcal{G}_{mix}$. This predicate serves as a precondition of the \mix\ and \swapmix\ actions.
    \item $\textup{same\_mixgraph}(s_1,s_2)$: if a given pair of qstates $s_1$ and $s_2$ belong to the same mix-graph, but are not connected by an edge. This enables the \swapmixaswap\ action between those qstates.
    \item $\textup{ps\_completed}(s)$: if the PS phase for a given qstate $s$ is finished.
\end{itemize}
\vspace{-1mm}

\noindent \emph{Actions:} there are 6 action templates representing the 6 architecture-specific operations described in Section~\ref{sec:qcc_graphcoloring}: \ps, \mix, \swap, \move, \swapps, and \swapmix. These actions act on the edges of the graph $\mathcal{G}_{machine}$ with appropriate qstates as action parameters. As outlined in Section~\ref{sec:qcc_graphcoloring}, the \swapmix\ action's $\textup{mixed}(s_1,s_2)$ effect conditions on whether or not the two involved qstates $s_1$ and $s_2$ are connected in the $\mathcal{G}_{mix}$ graph. Given that many temporal planners can not handle conditional effects, to allow us to use a wider range of planners, we compile it into two deterministic actions: (1) \swapmix: operates on two qstates $s_1$ and $s_2$ connected in $\mathcal{G}_{mix}$ (i.e., having $\textup{mixgraph\_edge}(s_1,s_2)$ as a condition) and combines both \mix\ and \swap\ action effects; and (2) \swapmixaswap: operates on two qstates $s_1$ and $s_2$ belonging to the same mix-graph but there is no edge connecting them (i.e., having $\textup{same\_mixgraph}(s_1,s_2) \wedge \neg \textup{mixgraph\_edge}(s_1,s_2)$ as its conditions) and has the same action duration as \swapmix\ while having the same effects as \swap. 
We also introduce an auxiliary instantaneous action \doneps$(s)$, which has a single effect $\textup{ps\_completed}(s)$, to specify that a qstate $s$ is done with the \ps\ phase and is ready to move on to the \mix\ phase. 
In our example shown in Figure~\ref{fig:example}, \doneps$(\psi_{1R})$ can be executed when $\textup{psed}(\psi_{1R},\psi_{2R})$ and $\textup{psed}(\psi_{1R},\psi_{4R})$ are achieved and its $\textup{ps\_completed}(\psi_{1R})$ effect in turn enables any \mix\ or \swapmix\ action involving $\psi_{1R}$, e.g. \mix($\psi_{1R},\psi_{1B}$) (see Figure~\ref{fig:PDDL}).

\vspace{1mm}

\noindent \emph{Initial state:} the initial state declares the initial values of all predicates: (1) $\textup{located\_at\_}q(s)$ and $\textup{empty\_}q$ specify the initial locations of all qstates $s$ (and if some physical qubits $q$ are empty); (2) $\textup{mixgraph\_edge}(s_1,s_2)$ and $\textup{same\_mixgraph}(s_1,s_2)$ values capture the $\mathcal{G}_{mix}$ graph; and (3) $\textup{edge}(s_1,s_2)$ values represent the connections in the input graph $\mathcal{G}$ to be colored.

\vspace{1mm}

\noindent \emph{Goal state:} the goal state specifies the list of \mix\ gates, represented by the $\textup{mixed}(s_1,s_2)$ predicates, that need to be achieved. The $\textup{ps\_completed}(s)$ conditions of the \mix\ and \swapmix\ actions (see Figure~\ref{fig:PDDL}) capture the ordering constraints between \ps\ and \mix\ gates and ensure that all requisite \ps\ gates will also be achieved.

%------
\begin{figure}[tb]
  \centering
  \includegraphics[width=\columnwidth]{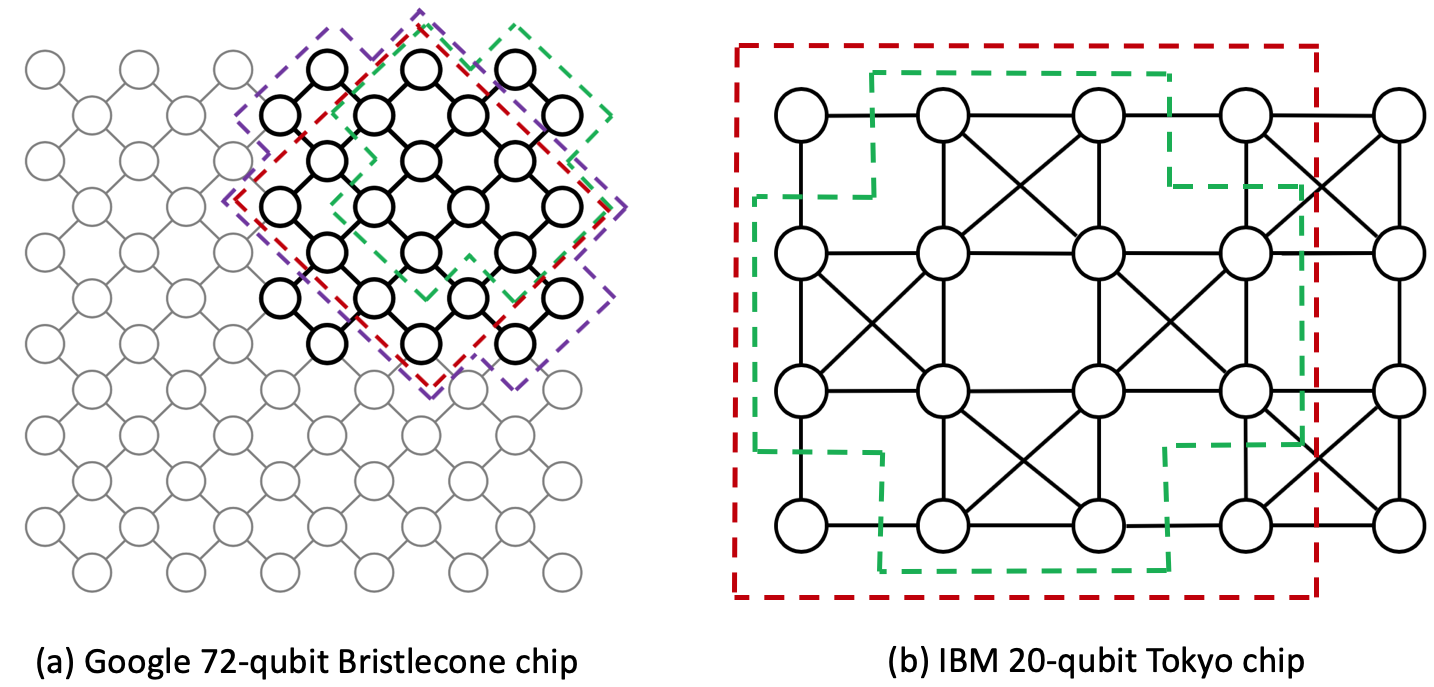}
  \caption{Google and IBM full chip configurations and the sections of 12 (green), 16 (red), 20 (purple), and 24 qubits used for our empirical evaluation.}
    \label{fig:chip_full}
\end{figure}
%-------

%-----------------
\section{Qubit Initialization}
\label{sec:initialization}

Qubit Initialization (QI) is the problem of assigning the initial locations of all qstates on the chip. Figure~\ref{fig:machine} shows examples of QI on the three machine configurations. Two approaches for QI were explored in the previous routing as Temporal Planning for MaxCut work~\cite{booth:icaps18}: (1) random initialization of qubits; and (2) for each qstate $s$, add an action \init($s,q$) to locate $s$ on the physical qubit $q$; all \init\ actions need to be carried out before any other action can start. 

Basically, the second approach combines QI with routing, resulting in the combined Routing-I problem~\cite{booth:icaps18}, with the hope that current state-of-the-art temporal planners would be able to find good initial locations for all qstates that support finding lower makespan plan. While ~\cite{booth:icaps18} compared different temporal planners on solving Routing-I for MaxCut, this work didn't report the difference in makespan between random initialization and Routing-I. Our current investigation for Routing as Temporal Planning for Graph Coloring shows that using a temporal planner to solve the Routing-I problems is not a clear-cut better option than random initialization (see results in Section~\ref{subsec:res_init}), and it doesn't perform well compared to careful manual initialization utilizing domain knowledge about the problem structure and the hardware chip layout. We believe this is due to: (1) the large number of possible initial configurations (for a $n$-qubit chip, the number of possible initial qstate allocations is $n!$, modulo potential symmetries); and (2) since initialization needs to be fully done before gate composition begins, it is difficult for the existing planners to compute a good heuristic estimate on which initial state is a good one to start from since they are the furthest states from the goals. The farther a given state from the goals, the harder to get a good heuristic estimate for that state.\\

%------
\begin{figure}[tb]
  \centering
  \includegraphics[width=0.8\columnwidth]{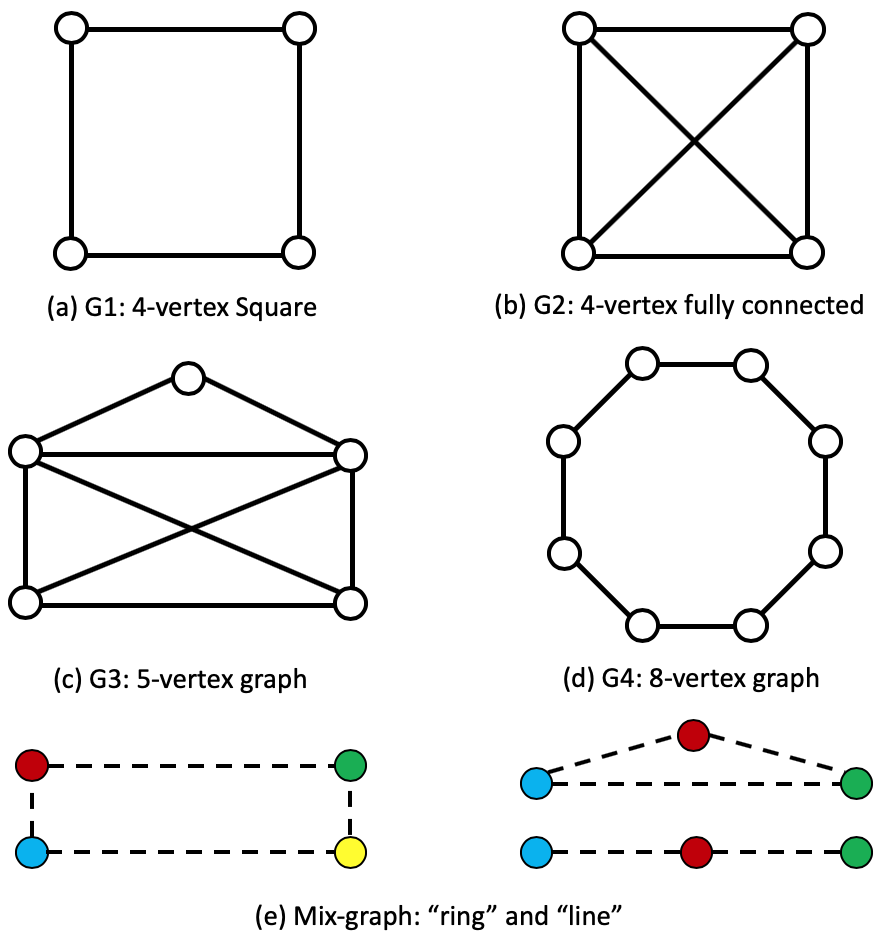}
  \caption{The benchmark instance set: Graphs to be colored (a-d) and mix-graph configurations (e).}
    \label{fig:exp_problems}
\end{figure}
%-------

\noindent {\bf Qubit initialization as classical planning:} instead of solving the combined Routing-I problem, we can heuristically solve the QI problem separately. Thus, we can try to find the initial locations $I$ of qstates so that the subsequent routing problem yields shorter makespan plans. As described in Section~\ref{sec:qcc_graphcoloring}, the routing for Graph Coloring requires each involved qstate to conduct all related \ps\ gates, and then all related \mix\ gates. Since the \ps\ gates will be done prior to the \mix\ gates, one heuristic is to initialize qstates so that if a pair of qstates $s_1$ and $s_2$ requires \ps($s_1, s_2$) as a goal, then they would be initially allocated on two qubits that are close to each other on the hardware chip. In our leading example, shown in Figure~\ref{fig:example} and ~\ref{fig:machine}, we would like to put the two qstates $\psi_{1R}$ and $\psi_{2R}$ close initially to minimize the total amount of time to execute the needed \swap\ gates to bring $\psi_{1R}$ and $\psi_{2R}$ to two connected qubits to enable the required \ps($\psi_{1R}, \psi_{2R}$) gate. To heuristically accomplish this objective, we setup and solve a cost-optimization classical planning problem $P'$, which is significantly simpler than the temporal planning for circuit routing problem $P$ described in Section~\ref{sec:qcc_temporalplanning}, as follows:

\vspace{1mm}

\noindent \emph{Predicates:} the following facts are represented by predicates for $P'$: $\textup{located\_at\_}q(s)$, $\textup{empty\_}q$, $\textup{edge}(s_1,s_2)$, and $\textup{psed}(s_1,s_2)$ (see Section~\ref{sec:qcc_temporalplanning} for their meaning). Predicates that are needed in $P$ but are not used in $P'$ are: $\textup{mixed}(s_1,s_2)$, $\textup{mixgraph\_edge}(s_1,s_2)$, $\textup{same\_mixgraph}(s_1,s_2)$, and $\textup{ps\_completed}(s)$.

\vspace{1mm}

\noindent \emph{Actions:} the {\em non-temporal} version of three actions \swap, \swapps, and \ps\ in the PDDL2.1 model for $P$ make up the action set for $P'$. The actions cost for \swap\ and \swapps\ in $P'$ are set to be equal to the duration of the temporal \swap\ and \swapps\ in $P$ while the action cost of \ps\ is set to a constant value 1.

\vspace{1mm}

\noindent \emph{Initial State:} declares the initial locations of all qstates $s$ by setting up the values of $\textup{located\_at\_}q(s)$ and $\textup{empty\_}q$; values of $\textup{edge}(s_1,s_2)$ represent the connections in the input graph $\mathcal{G}$ to be colored.

\vspace{1mm}

\noindent \emph{Goal State:} specifies the list of \ps\ gates that need to be achieved.

\vspace{1mm}

We then use a classical planner to find a solution $\pi$ for $P'$ with the objective function of minimizing the total plan cost. The final locations of all qstates \emph{after} executing $\pi$ are then used as the initial qstate allocation for the original temporal planning problem $P$. Since the goal of $P'$ is to achieve all \ps\ gates in $P$, we conjecture that qstate pairs $(s_1,s_2)$ that are engaged in the \ps$(s_1,s_2)$ gates in $P$ would likely be close to each other when $\pi$ is done executing. If either $s_1$ or $s_2$ needs to be moved after \ps$(s_1,s_2)$ is accomplished to accommodate another \ps\ gate (e.g., \ps$(s_1,s_3)$), then the objective function of mimimizing the total action cost will enforce a sequence of \swap\ and \swapps\ gates that take the minimum total time to accomplish it. Thus, the final qstate locations after $\pi$ is done executing setup an attractive initial locations for the original planning problem $P$ where the same set of \ps\ gates need to be accomplished before the subsequent \mix\ gates.

Given the simplified problem $P'$ and the larger number of classical planners available, compared to temporal planners, the QI-as-classical-planning problem should be significantly easier to solve compared to the original Routing as Temporal Planning problem.

%----------
\section{Empirical Evaluation}
\label{sec:evaluation}

In this section, we will present preliminary empirical evaluation of different temporal planners applied to routing for Graph Coloring.

\vspace{1mm}

\noindent {\bf Hardware Architecture:} We use the recently published hardware chip architecture layouts from Rigetti (Aspen), Google (Bristlecone), and IBM (Tokyo). The full-size chip from Rigetti is shown in Figure~\ref{fig:machine}(a), and sections of the full chip from Google and IBM are shown in Figure~\ref{fig:machine}(b) and (c) respectively. Figure~\ref{fig:chip_full} shows the full 72-qubit Bristlecone and the 20-qubit Tokyo architectures. Overall, the hardware architectures we use have 12--24 qubits:
(1) Rigetti: 12-qubit section and the full 16-qubit chip (Figure~\ref{fig:machine}(a));
(2) Google: 12-, 16-, 20- and 24-qubit sections of the Bristlecone 72-qubit architecture (Figure~\ref{fig:chip_full}(a));
(3) IBM:\@ 12- and 16-qubit sections and the full 20-qubit Tokyo chip (Figure~\ref{fig:chip_full}(b)).

\vspace{1mm}

\noindent {\bf Software:} Previous work on routing for MaxCut used TFD~\cite{TFD}, LPG~\cite{LPG}\footnote{Since LPG is a stochastic planner, for each problem, we ran LPG 10 times and report its best result.}, CPT~\cite{vidal:aij2006}, POPF~\cite{popf}, and SGPlan~\cite{SGPlan2}\footnote{We are considering TSGP as an alternative to SGPlan.}. We exclude CPT due to poor scalability (for both Graph Coloring and Max Cut) and replace POPF with the more modern code of another planner, OPTIC~\cite{optic}, in the same family. All planners were run in the anytime setting (except SGPlan which does not have this mode). For solving the QI-as-planning problem, we use the classical planner Fast Downward~\cite{FastDownward} with the LAMA 2011 ~\cite{lama2011} configuration.

\vspace{1mm}

\noindent {\bf Problem specifications:} We test our approach on a range of randomly selected Graph Coloring instances with 4, 5, and 8 vertices and either the ``line'' or ``ring'' mix-graph.
They are shown in Figure~\ref{fig:exp_problems}(a--d) as graphs G1--4. For example, graph G2R4 (Table~\ref{tab:result1}, ~\ref{tab:analytical_bounds}, and ~\ref{tab:qi_compare}) means solving graph G2 in Figure~\ref{fig:exp_problems} with the ``Ring" mix-graph with 4 colors.
The number of vertices $n$ in the graph and the number of colors $k$ in the mix-graph will require the number of physical qubits $n \cdot k$ within the range of 12-24 qubits, as described in the previous paragraph.

\vspace{1mm}

\noindent {\bf Gate durations:} For all hardware architectures and for all edges in the hardware chips, the following gate durations are used: dur(\swap) = 4, dur(\move) = 4, dur(\ps) = 3, dur(\swapps) = 4, dur(\mix) = 1, and dur(\swapmix) = 1. These are effective durations based on logical gate synthesis using a common native gate sets, e.g., those available on Rigetti's chips~\cite{XYrigetti}. Results were collected on a RedHat Linux VM running on a Macbook Pro with 4GB of RAM. The runtime limit was set to 600 seconds for problems involving 12--16 qubits and 1200 seconds for problems involving 20--24 qubits.

%----------
\begin{table}
\begin{center}
\captionsetup{justification=centering}
\caption{Plan quality comparison in solving problems shown in Figure~\ref{fig:exp_problems} with different hardware architectures shown in Figure~\ref{fig:machine} and ~\ref{fig:chip_full}.
Qubit initialization is done manually using expert knowledge, similar to those in Figure~\ref{fig:machine}.
\textbf{bold} values indicate the best overall makespan.
``-'' indicates either the planner ran out of time before finding a solution (OPTIC) or crashed (SGPlan).
}\label{tab:result1}
\begin{tabular}{|l|c||c|c|c|c|}
\hline
                            & Graph & TFD & OPTIC & LPG & SGPLAN \\ \hline \hline
\multirow{2}{*}{Rigetti-12} & G1R3  &  {\bf 28} &  {\bf 28}   &  31 &   61   \\ \cline{2-6} 
                            & G1L3  &  {\bf 27} &  42   &  33 &   44   \\ \hline
\multirow{2}{*}{Google-12}  & G1R3  &  {\bf 22} &  45   &  46 &   83   \\ \cline{2-6} 
                            & G1L3  &  {\bf 21} &  38   &  41 &   68   \\ \hline
\multirow{2}{*}{IBM-12}     & G1R3  & {\bf 23}  &  37   &  31 &   51   \\ \cline{2-6} 
                            & G1L3  & {\bf 17}  &  30   &  36 &  39    \\ \hline
\multirow{2}{*}{Rigetti-16} & G1R4  & {\bf 31}  &   -   & 80  &  94    \\ \cline{2-6} 
                            & G2R4  & {\bf 74}  &   -   & 119 &  156   \\ \hline
\multirow{2}{*}{Google-16}  & G1R4  & {\bf 19}  &  46   & 20  &  34    \\ \cline{2-6} 
                            & G2R4  & 76  &  49   & {\bf 37}  &  48    \\ \hline
\multirow{2}{*}{IBM-16}     & G1R4  & 43  &  -    & 26  &  {\bf 20}    \\ \cline{2-6} 
                            & G2R4  & 79  &  58   & 49  &  {\bf 29}    \\ \hline
Google-20                   & G3R4  & {\bf 64}  &  -    & 86  & 106    \\ \hline
IBM-20                      & G3R4  & 113 &  -    & {\bf 94}  &   -    \\ \hline
Google-24                   & G4R3  & 125 &  {\bf 64}   & 83  &   -    \\ \hline
\end{tabular}
\end{center}
\end{table}
%---------

%-----------------
\begin{table}
\begin{center}
\captionsetup{justification=centering}
\caption{Comparing against analytical bounds for special hardware architectures: grid and line. Highlight in {\bf bold} are makespan values that are better than the analytical bound. \uline{Underlined} values are the best makespan produced by any planner when this best makespan is worse than the analytical bound.}\label{tab:analytical_bounds}
\begin{tabular}{|l||c|c|c|c|c|}
\hline
                  & Analyt. Bound & TFD & OPTIC & LPG & SGPLAN \\ \hline \hline
4 $\times$ 3 Grid &     19       & 20  &   35  & {\bf 16}  &  {\bf 13}    \\ \hline
4 $\times$ 4 Grid &     20       & 30  &   56  & {\bf 20}  &  38    \\ \hline
5 $\times$ 4 Grid &     24       & 54  &   116 & 79  &  \uline{46}    \\ \hline
8 $\times$ 3 Grid &     35       & {\bf 25}  &   53   & 36  &   -    \\ \hline
4 $\times$ 3 Line &     64       & {\bf 51}  &   77  & {\bf 48}  &  90    \\ \hline
4 $\times$ 4 Line &     80       & {\bf 71}  &   116 & 107 &  177   \\ \hline
5 $\times$ 4 Line &     100      & \uline{118} &   -   & 140 &  194   \\ \hline
8 $\times$ 3 Line &     128      & {\bf 81}  &   -   & 157 &  267   \\ \hline
\end{tabular}
\end{center}
\end{table}
%------------------

\vspace{-3mm}

%---------
\subsection{Solving Circuit Routing with Temporal Planner}
\label{subsec:res_solving}

Table~\ref{tab:result1} compares results between 4 temporal planners. Overall, TFD and LPG are the only two planners that can solve all problems within the time limit. In terms of solution quality, TFD is most often the best planner, especially for smaller problems. However, each planner has at least one case where it performs best. SGPlan generally returns the worst quality plan but it also performs the best in the two problems on IBM 16-qubit architecture. With regard to \emph{ring} vs \emph{line} mix-graph, as expected, the ring version of the 12-qubit problems normally has slightly longer makespan value than the line counterpart, given that it requires additional \mix\ goal gates. Between different hardware architectures, the Rigetti machine has many fewer connections for the same chip size (12 or 16-qubit), which leads to fewer parallel routes to move qstates. This led to longer makespan plans, in general, compared to solving the same problems on the Google and IBM architectures. Comparing the Google and IBM architectures of the same size, we expect planners should be able to find equal or shorter makespan plans given that they have the same overall shape except that the IBM ones have some additional connections. While that generally holds true for the 12-qubit version, looking at the results for the 16-qubit version we see TFD, OPTIC, and LPG perform worse on the IBM architecture than the Google architecture. It seems that the additional connections enlarge the planning search space and confuse the planners: they are not able to exploit the additional connectivity to find better quality plans. SGPlan is the only planner that show marked better performance on the IBM 16-qubit architecture.

%-----------------
\begin{table}
\begin{center}
\captionsetup{justification=centering}
{\caption{Improvement in makespan when solving a problem using qubit initialization provided by a classical planner (Section~\ref{sec:initialization}). Example: entry 8.1\%(8/10) for cell IBM-16, G2R4, OPTIC means: among 10 randomly generated QIs for solving G2R4 on IBM-16, OPTIC can solve 8 with random QI while it can solve all 10 with QIs produced by the Fast Downward classical planner solving the same random problems. Accross the 8 problems that are solved with both QI setups, the average makespan improvement is 8.1\%. When the number of solved problem is not listed, it means all 10 are solved with both QI options. \textcolor{red}{RED} indicates entries where the Fast Downward's QI performs worse than random QI.}\label{tab:qi_result1}}
\begin{tabular}{|l|c||c|c|c|}
\hline
                            & Graph & TFD & OPTIC & SGPlan \\ \hline \hline
\multirow{2}{*}{Rigetti-12} & G1R3  & 17.9\%  &  14.4\%   &  19.1\%     \\ \cline{2-5} 
                            & G1L3  & 10.5\%  &  2.7\%   &   3.52\%  \\ \hline
\multirow{2}{*}{Google-12}  & G1R3  & 9.7\%  &  7.9\%   &   \textcolor{red}{-5.7}\%    \\ \cline{2-5} 
                            & G1L3  & 16.3\%  &  12.3\%   &   13.1\%   \\ \hline
\multirow{2}{*}{IBM-12}     & G1R3  & 9.2\%  &  2.1\%   &   10.9\%  \\ \cline{2-5} 
                            & G1L3  & 10.2\%  &  18.4\%   &   17.4\%  \\ \hline
\multirow{2}{*}{Rigetti-16} & G1R4  & 13.6\%  &  23.8\%(7/10)   &  15.7\%   \\ \cline{2-5} 
                            & G2R4  & 20.4\%  &  24.6\%(8/10)   &   18.1\%   \\ \hline
\multirow{2}{*}{Google-16}  & G1R4  & 7.4\%  &  5.3\%   &  12.3\%   \\ \cline{2-5} 
                            & G2R4  & 17.5\% (\textcolor{red}{10/9})  &  7.9\%(9/10)   &  11.3\%   \\ \hline
\multirow{2}{*}{IBM-16}     & G1R4  & 11.2\%  &  \textcolor{red}{-3.9}\%   &  14.0\%   \\ \cline{2-5} 
                            & G2R4  & 6.4\% (\textcolor{red}{7/6})  &  8.1\%(8/10)   &   7.2\%   \\ \hline
Google-20                   & G3R4  & 23.0\%  &  2.6\%(4/9)   &   23.7\%   \\ \hline
IBM-20                      & G3R4  & 12.6\%  &  (2/3)   &     - \\ \hline
Google-24                   & G4R3  & 6.8\% (9/9) &  (0/2)   &    - \\ \hline
\end{tabular}
\end{center}
\end{table}
%-----------------

\vspace{-4mm}

%-----------------
\begin{table*}
\begin{center}
\begin{footnotesize}
\captionsetup{justification=centering}
{\caption{Comparing different initialization strategies. \textbf{M}: \textbf{M}anual initialization (results from Table~\ref{tab:result1}); \textbf{I}: Solving the combined QI + Routing (Routing-I) problem in one temporal-planning run (see Section~\ref{sec:initialization}); \textbf{Random} and Fast Downward(\textbf{FD}) are initialization setups explained in Table~\ref{tab:qi_result1} with ``AVG'' and ``Best'' representing the average and best values across 10 random QIs. Values in \textcolor{red}{RED} show the best value across all setups; values in \textbf{bold} are best for a given planner. Random vs FD qubit initialization: Light Yellow background indicates better makespan by FD while Light Cyan indicates better makespan for random QI.
}\label{tab:qi_compare}}
\setlength\tabcolsep{2.8pt} % default value: 6pt
\begin{tabular}{|l|l||c|c|c|c|c|c||c|c|c|c|c|c||c|c|c|c|c|c||c|}
\hline
\multicolumn{2}{|l|}{} & \multicolumn{6}{|c|}{TFD} & \multicolumn{6}{|c|}{OPTIC} & \multicolumn{6}{|c|}{SGPlan} & LPG \\ \hline
\multicolumn{2}{|c|}{Problem} & M & I & \begin{tabular}[c]{@{}c@{}}Rand.\\ (AVG)\end{tabular} & \begin{tabular}[c]{@{}c@{}}Rand.\\ (Best)\end{tabular} & \begin{tabular}[c]{@{}c@{}}FD\\ (AVG)\end{tabular} & \begin{tabular}[c]{@{}c@{}}FD\\ (Best)\end{tabular} & M & I & \begin{tabular}[c]{@{}c@{}}Rand.\\ (AVG)\end{tabular} & \begin{tabular}[c]{@{}c@{}}Rand.\\ (Best)\end{tabular} & \begin{tabular}[c]{@{}c@{}}FD\\ (AVG)\end{tabular} & \begin{tabular}[c]{@{}c@{}}FD\\ (Best)\end{tabular} & M & I & \begin{tabular}[c]{@{}c@{}}Rand.\\ (AVG)\end{tabular} & \begin{tabular}[c]{@{}c@{}}Rand.\\ (Best)\end{tabular} & \begin{tabular}[c]{@{}c@{}}FD\\ (AVG)\end{tabular} & \begin{tabular}[c]{@{}c@{}}FD\\ (Best)\end{tabular} & M \\ \hline \hline
\multirow{2}{*}{Rigetti-12} & G1R3 & \textcolor{red}{28} & - & 61.2 & 53 & 50.1 & \cellcolor{lightyellow}42 & \textcolor{red}{28} & 76 & 73.7 & 63 & 62.7 & \cellcolor{lightyellow}44 & 61 & 89 & 94 & 69 & 75.9 & \cellcolor{lightyellow}\textbf{53} & 31 \\ \cline{2-21} 
 & G1L3 & \textcolor{red}{27} & - & 52.3 & 44 & 46.3 & \cellcolor{lightyellow}39 & \textbf{42} & 62 & 65 & 49 & 62.9 & \cellcolor{lightyellow}48 & \textbf{44} & 84 & 73.4 & 62 & 70.5 & \cellcolor{lightyellow}59 & 33\\ \hline
\multirow{2}{*}{Google-12} & G1R3 & \textcolor{red}{22} & 47 & 39.8 & 33 & 35.8 & \cellcolor{lightyellow}27 & 45 & 40 & 48.7 & \cellcolor{lightcyan}\textbf{31} & 43.7 & 34 & 83 & 77 & 60.9 & \cellcolor{lightcyan}\textbf{43} & 65.4 & 52 &  46 \\ \cline{2-21} 
 & G1L3 & \textcolor{red}{21} & 50 & 38.4 & 34 & 31.9 & \cellcolor{lightyellow}25 & 38 & 28 & 43 & 34 & 37.7 & \cellcolor{lightyellow}\textbf{27} & 68 & 69 & 62.1 & 46 & 52 & \cellcolor{lightyellow}\textbf{38} & 41 \\ \hline
\multirow{2}{*}{IBM-12} & G1R3 & \textcolor{red}{23} & 35 & 33.8 & 27 & 30.5 & \cellcolor{lightyellow}26 & \textbf{37} & 61 & 47.5 & 40 & 45.5 & \cellcolor{lightyellow}\textbf{37} & 51 & 63 & 56.5 & 47 & 49.8 & \cellcolor{lightyellow}\textbf{43} & 31 \\ \cline{2-21} 
 & G1L3 & \textcolor{red}{17} & 25 & 29.9 & \cellcolor{lightcyan}19 & 26.4 & 21 & 30 & 39 & 44.2 & 31 & 36.1 & \cellcolor{lightyellow}\textbf{25} & 39 & 50 & 52.9 & 42 & 43.2 & \cellcolor{lightyellow}\textbf{38} & 36 \\ \hline
\multirow{2}{*}{Rigetti-16} & G1R4 & \textcolor{red}{31} & 77 & 62.4 & 49 & 51.7 & \cellcolor{lightyellow}44 & - & 66 & 78.4 & 52 & 58.3 & \cellcolor{lightyellow}\textbf{43} & 94 & 118 & 119.3 & 92 & 97.4 & \cellcolor{lightyellow}\textbf{68} &  80 \\ \cline{2-21} 
 & G2R4 & 74 & - & 83.2 & 65 & 64.5 & \cellcolor{lightyellow}\textcolor{red}{56} & - & 85 & 94.5 & 76 & 72.1 & \cellcolor{lightyellow}\textbf{63} & 156 & 140 & 139 & 105 & 111 & \cellcolor{lightyellow}\textbf{89} & 119 \\ \hline
\multirow{2}{*}{Google-16} & G1R4 & \textcolor{red}{19} & - & 42.4 & 36 & 39.2 & \cellcolor{lightyellow}31 & 46 & 51 & 58.9 & 47 & 55.8 & \cellcolor{lightyellow}\textbf{39} & \textbf{34} & 94 & 86.3 & 57 & 73.5 & \cellcolor{lightyellow}\cellcolor{lightyellow}55 & 20 \\ \cline{2-21} 
 & G2R4 & 76 & - & 98.3 & 55 & 79.8 & \cellcolor{lightyellow}\textbf{43} & 49 & 68 & 71.8 & 61 & 65.7 & \cellcolor{lightyellow}\textbf{48} & \textbf{48} & 103 & 112.3 & 83 & 98.1 & \cellcolor{lightyellow}75 & \textcolor{red}{37}  \\ \hline
\multirow{2}{*}{IBM-16} & G1R4 & 43 & - & 60.9 & 47 & 53.7 & \cellcolor{lightyellow}\textbf{31} & - & 57 & 50.8 & 34 & 54.1 & \cellcolor{lightyellow}\textbf{32} & \textcolor{red}{20} & - & 77.3 & 63 & 65.9 & \cellcolor{lightyellow}53 & 26 \\ \cline{2-21} 
 & G2R4 & 79 & - & 85 & 74 & 79.5 & \cellcolor{lightyellow}\textbf{70} & 58 & 56 & 73.5 & 57 & 64.6 & \cellcolor{lightyellow}\textbf{51} & \textcolor{red}{29} & - & 89.2 & 67 & 82.7 & \cellcolor{lightyellow}52 & 49 \\ \hline
Google-20 & G3R4 & 64 & - & 128.9 & 64 & 90.7 & \cellcolor{lightyellow}\textcolor{red}{58} & - & - & 86 & 76 & 82.9 & \cellcolor{lightyellow}\textbf{62} & 106 & - & 152.3 & 115 & 115.1 & \cellcolor{lightyellow}\textbf{88} & 86 \\ \hline
IBM-20 & G4R4 & 113 & - & 111.7 & 82 & 96.3 & \cellcolor{lightyellow}\textcolor{red}{48} & - & - & 77.5 & \cellcolor{lightcyan}\textbf{70} & 81 & 76 & - & - & - & - & - & - & 94\\ \hline
Google-24 & G4R3 & 125 & - & 155.7 & 134 & 143.6 & \cellcolor{lightyellow}\textbf{113} & 64 & \textcolor{red}{37} & - & - & 72.5 & \cellcolor{lightyellow}61 & - & - & - & - & - & - & 83 \\ \hline
\end{tabular}
\end{footnotesize}
\end{center}
\end{table*}
%------------------

%------------------
\subsection{Planner vs Analytical Bounds}
\label{subsec:analytical_bound}

 Without another systematic approach to solve Routing for Graph Coloring to compare with, one question remains: \emph{how good is the quality of solutions returned by temporal planners in this domain?}
 For comparison, we can use simple manually constructed solutions, which are currently available for certain problem setups, as an upper bound~\cite{ogorman2019generalized}. Recall that for coloring a graph of $n$ vertices with $k$ colors on a hardware chip, we use $n \cdot k$ qubits. For hardware that supports the gates described in Section~\ref{sec:qcc_graphcoloring}, we can get valid plans for the two hardware architectures: an $n \cdot k$ ``grid'' layout and a $n \cdot k$-length ``line'' layout with the following makespans:
 
 \vspace{1mm}
 
 \noindent $\mathrm{makespan}_{LINE} \leq n \cdot \tau_{\mathrm{SWAP\_PS}} + nk \cdot \tau_{\mathrm{SWAP}} + k \cdot \tau_{\mathrm{SWAP\_MIX}}$\\
 \noindent $\mathrm{makespan}_{GRID} \leq n \cdot \tau_{\mathrm{SWAP\_PS}}  + k \cdot \tau_{\mathrm{SWAP\_MIX}}$\\

This can be accomplished through a predetermined qubit initialization and operation sequence such that while there will be \swap\ gates required for the ``line" layout, the ``grid" layout only needs combined gates (i.e., \swapmix\ and \swapps) to accomplish all goal gates. Table~\ref{tab:analytical_bounds} shows that while each planner's performance is quite inconsistent, the best plan returned by the planning approach compare well with reasonable upper bounds for those ``grid" and ``line" chip layout.

%---------
\subsection{Qubit Initialization as Classical Planning}
\label{subsec:res_init}

To measure the effect of qubit initialization for Graph Coloring by solving a classical planning problem, for each of the 15 problem configurations described in the previous Section~\ref{subsec:res_solving} and shown in Table~\ref{tab:result1}, we: (1) generate 10 random qubit initializations; (2) use the classical planner Fast Downward to solve the qubit initialization problem as described in Section~\ref{sec:initialization} (time limit: 200 seconds); (3) solve two versions of the problem with: (i) random initialization and (ii) with initialization given by the solution returned by Fast Downward. Table~\ref{tab:qi_result1} shows the makespan improvement of using the same planner\footnote{We use the same running time limit as in collecting results for Table~\ref{tab:result1}}, solving (ii) over solving (i); averaged over 10 random problem for each configuration. We exclude LPG from this experiment due to its randomness nature. LPG would employ different random seeds when solving (i) and (ii), making the comparison not justifiable.

In essence, Table~\ref{tab:qi_result1} shows that given a (random) qubit initialization, different planners benefit from running a classical planner such as Fast Downward as a pre-processing phase to potentially find better initial locations for all qstates. The results show that all three planners benefit from this step for a majority of testing scenarios: solving higher overall number of instances while experiencing makespan improvement with the qubit initialization calculated by Fast Downward. There are only two cases in which average makespan increases: OPTIC/ IBM-16/G1R4 and SGPlan/Google-12/G1R3. The reason is likely that the majority of the random initializations happen to be very good starting points and Fast Downward's plans move qstates to overall worse initializations. Note that the classical planning setup for QI approximates and relaxes the problem by: (1) ignoring the \mix\ goals; (2) removing the temporal aspects (and thus finding a sequential non-parallel plan). Therefore, it's possible that the final qstate locations provided by the Fast Downward solution is worse than the starting random initialization.\\

\noindent {\bf Overall Comparison:} Table~\ref{tab:qi_compare} shows the makespan comparison between different settings for all planners. Overall, TFD consistently provides the best plan quality with manual initialization for smaller problems. For larger problems, where finding a good manual initialization is likely harder, different planners return the best quality solutions for different problem settings. Taking each individual planner, the data show, as expected, the best results are from either manual initialization or provided by using the Fast Downward (FD) planner. Between those two settings, FD provides better results for TFD and OPTIC on larger problems; for SGPlan, manual initialization provides better results for larger problems. While Routing-I, where a temporal planner solves both the QI and routing problem in one planning model, would be the most convenient setup, the results show that existing temporal planners are still not able to solve that problem effectively. Only one best makespan is found using this problem setup, provided by the OPTIC planner. It's also worth pointing out that while TFD is the most promising planner for routing, it performs worst on Routing-I with only 5/15 problems solved. Consistent with the results shown in Table~\ref{tab:qi_result1}, using the QI result provided by the Fast Downward planner, all three planners produce smaller makespan values (both average and best) compared to the random QI counterpart.

In summary, at large scale and for general instances, it is infeasible to manually solve the QI problem; for those cases, we show that the initialization process can be automated using a classical planner. The most promising planner is TFD and it should be the first one to try out. However, given that SGPlan is by far the fastest among all 4 tested planners, solving all tested problems within a few seconds, it's also worth running it besides other planners. While SGPlan is very inconsistent, it sometimes (see Table~\ref{tab:qi_compare} and~\ref{tab:analytical_bounds}) produces plans with makespan values much lower than other planners.

%-------------
\section{Conclusion and Future Work}
\label{sec:conclusion}

In this paper, we describe our recent investigation into using model-based planning technology to solve the quantum circuit routing for QAOA applied to the Graph Coloring problem: temporal planner for solving the Routing and classical planner for solving the QI problem. Our empirical evaluation shows that temporal planners can solve problems with diverse setups effectively and compare reasonably to the best known analytical bounds on special cases.
Qubit initialization as classical planning, utilising the Fast Downward planner, provides makespan improvement in the majority of problem configurations and provides an attractive alternative to manual qubit initialization using domain expert knowlege.

There are several directions we are pursuing in future work. First, we are working on running some plans produced by the temporal planners described in this paper on actual, physical hardware chips. Second, besides strengthening our current approach of using classical planning for qubit initialization with alternative PDDL encodings, some go beyond accomplishing just the PS phase by combining both the PS and MIX phases, we are also experimenting with several other approaches to QI such as posing it as a Quadratic Assignment(QA)~\cite{qa:burkard} problem, which can then be solved using a quadratic programming solver such as CPLEX. While the initial result show poor scaling for the quadratic programming formulation, some heuristic approaches for solving the QA problem look promising. Third, we would like to analyze better the synergy between different temporal planning algorithms and the problem structures that are most suitable to them. Lastly, several portfolio approaches have been showing good performances at the recent International Planning Competition (IPC); we would like to investigate their performance on the Routing for Graph Coloring domain, and compare them to our current set of temporal planners used in this paper.

\bibliography{ecai}
\end{document}